\documentclass[conference]{IEEEtran}
\IEEEoverridecommandlockouts
\usepackage[english]{babel}
\usepackage{amsthm}
\usepackage{titlesec}
\usepackage[ruled,vlined]{algorithm2e}
\usepackage{algcompatible}
\usepackage{graphicx}
\usepackage{float}
\usepackage{caption}
\usepackage{tikz}
\usepackage{changepage}
\usepackage[utf8]{inputenc}
\usepackage{pgfplots} 
\usepackage{pgfgantt}
\usepackage{pdflscape}
\usepackage[export]{adjustbox}
\pgfplotsset{compat=newest} 
\pgfplotsset{plot coordinates/math parser=false}
\pgfplotsset{compat=1.18}
\usepackage[justification=centering]{caption}
\usepackage{pgfplots}
\usetikzlibrary{spy}

\newtheorem{innercustomgeneric}{\customgenericname}
\providecommand{\customgenericname}{}
\newcommand{\newcustomtheorem}[2]{%
  \newenvironment{#1}[1]
  {%
   \renewcommand\customgenericname{#2}%
   \renewcommand\theinnercustomgeneric{##1}%
   \innercustomgeneric
  }
  {\endinnercustomgeneric}
}

\newcustomtheorem{customthm}{Theorem}
\newcustomtheorem{customlemma}{Lemma}

\usepackage{amsmath}
\usepackage{acronym}
\usepackage{amssymb}
\usepackage{mathtools}
\usepackage{url}
\usepackage{graphicx}  
\usepackage{float}  

\newcommand{\yy}{\mathbf{y}}
\newcommand{\xx}{\mathbf{x}}

\newcommand{\bb}{\mathbf{b}}
\newcommand{\hh}{\mathbf{h}}

\newcommand{\pp}{\mathbf{p}}

\newcommand{\BB}{\mathbf{B}}
\newcommand{\UU}{\mathbf{U}}

\newcommand{\HH}{\mathbf{H}}
\newcommand{\ff}{\mathbf{f}}

\newcommand{\WW}{\mathbf{W}}
\newcommand{\QQ}{\mathbf{Q}}
\newcommand{\FF}{\mathbf{F}}

\newcommand{\XX}{\mathbf{X}}
\newcommand{\RR}{\mathbf{R}}

\newcommand{\Tr}{\text{Tr}}
\newcommand{\diag}{\text{diag}}

\newcommand{\hermit}{\mathsf{H}}

\usepackage{accents}

\newcommand{\TT}{\mathbf{T}}

\newcommand{\MM}{\mathbf{M}}

\newcommand{\AAb}{\mathbf{A}}

\newcommand{\norm}[1]{\left\lVert#1\right\rVert}

\newcommand{\TTheta}{\boldsymbol{\Theta}}
\newcommand{\ttheta}{\boldsymbol{\theta}}
\newcommand{\ggc}{\mathbf{g}}
\newcommand{\vecc}{\text{vec}}

\newcommand{\GGc}{\mathbf{G}}

\acrodef{SISO}[SISO]{single-input single-output}
\acrodef{AP}[AP]{access point}
\acrodef{UE}[UE]{user equipment}
\acrodef{ULA}[ULA]{uniform linear array}
\acrodef{CPU}[CPU]{central processing unit}
\acrodef{FPP}[FPP]{Feasible-point pursuit}
\acrodef{LoS}[LoS]{line of sight}
\acrodef{NLoS}[NLoS]{non line of sight}
\acrodef{RCS}[RCS]{radar cross section}
\acrodef{AoD}[AoD]{angle of departure}
\acrodef{AoA}[AoA]{angle of arrival}
\acrodef{CRB}[CRB]{Cramer-Rao bound}
\acrodef{FIM}[FIM]{Fisher information matrix}
\acrodef{AN}[AN]{artificial noise}
\acrodef{SINR}[SINR]{signal-to-interference-plus-noise ratio}
\acrodef{SNR}[SNR]{signal-to-noise ratio}
\acrodef{QoS}[QoS]{quality of service}
\acrodef{SDR}[SDR]{semi-definite relaxation}
\acrodef{SDP}[SDP]{semi-definite relaxation}
\acrodef{ISAC}[ISAC]{integrated sensing and communications}
\acrodef{PLS}[PLS]{physical layer security}
\acrodef{CSI}[CSI]{channel state information}
\acrodef{MUI}[MUI]{multi-user interference}
\acrodef{RIS}[RIS]{Reconfigurable intelligent surface}
\acrodef{AO}[AO]{alternating optimization}
\acrodef{SIMO}[SIMO]{Single Input Multiple Output}
\acrodef{MISO}[MISO]{multiple-intput single output}
\acrodef{MIMO}[MIMO]{multiple-input multiple-output}
\acrodef{MU}{multi-user}
\acrodef{BS}[BS]{base station}
\acrodef{CEE}[CEE]{channel estimation error}
\acrodef{CCP}[CCP]{convex-concave procedure}
\acrodef{MRT}[MRT]{maximum-ratio transmission}
\acrodef{MM}[MM]{Minorization-Maximization}
\acrodef{PSD}[PSD]{positive semi-definite}
\begin{document}

\title{ Destructive and Constructive RIS Beamforming in an ISAC Multi-User MIMO Network}

\author{Steven~Rivetti$^\dagger$,
Özlem~Tu$\Breve{\text{g}}$fe~Demir$^*$,
Emil~Björnson$^\dagger$, 
        Mikael~Skoglund$^\dagger$ \\
        
       {\small$^\dagger$School of Electrical Engineering and Computer Science (EECS),
        KTH Royal Institute of Technology, Sweden} \\
        
        {\small$^*$Department of Electrical-Electronics Engineering, TOBB University of Economics and Technology, Ankara, Türkiye}
        
       \thanks{
This work was supported by the SUCCESS project (FUS21-0026), funded by the Swedish Foundation for Strategic Research.}}

\maketitle

\begin{abstract}
Integrated sensing and communication (ISAC) has already established itself as a promising solution to the spectrum scarcity problem, even more so when paired with a reconfigurable intelligent surface (RIS), as RISs can shape the propagation environment by adjusting their phase-shift coefficients.
Albeit the potential performance gain, a RIS is also a potential security threat to the system. In this paper, we explore both the positive and negative sides of having a RIS  in a multi-user multiple-input multiple-output (MIMO) ISAC network. We first develop an alternating optimization algorithm, obtaining the active and passive beamforming vectors that maximize the sensing signal-to-noise ratio (SNR) under minimum signal-to-interference-plus-noise ratio (SINR) constraints for the communication users and finite power budget.
We also investigate the destructive potential of the RIS by devising a RIS phase-shift optimization algorithm that minimizes the sensing SNR while preserving the same minimum communication SINR previously guaranteed by the system.
We further investigate the impact of the RIS's individual element failures on the system performance.
The simulation results show that the RIS performance-boosting potential is as good as its destructive one and that both of our optimization strategies are hindered by the investigated impairments.

\end{abstract}

\begin{IEEEkeywords}
RIS, malicious RIS, ISAC, MIMO, RIS element failures
\end{IEEEkeywords}

\section{Introduction}
\acp{RIS} have captured significant interest from both academia and industry for their ability to control challenging radio propagation environments intelligently. By using low-cost passive reflecting elements that induce a controllable phase-shift to incoming waveforms, RISs can effectively steer reflected signals towards desired locations while reducing interference at undesired points \cite{alexandropoulos2021reconfigurable,chen2023simultaneous}. Moreover, RISs have recently emerged as a promising technology in \ac{ISAC} applications, increasing the available spatial degrees of freedom. 
In \cite{liu2022joint}, the sum rate of the communication \acp{UE} is maximized under the worst-case sensing \ac{SNR} constraint.
On the other hand, \cite{zhong2023joint} defines a sensing \ac{SINR} by taking into account the presence of interfering objects along the sensing path: this sensing SINR is then maximized in a weighted fashion alongside minimizing the multi-user interference.

Despite their cost-effective deployment and potential to enhance wireless communication and sensing performance, RISs present critical \ac{PLS} challenges. In particular, RISs could potentially be exploited as malicious entities that instead compromise communication and sensing performance. \cite{wang2022wireless} gives an overview of the possible ways in which an illegally deployed RIS can hinder system performance, characterizing signal leakage from the legitimate system to the said RIS and the potential damage of interference attacks launched by the former.
Adding artificial noise onto the legitimate system's transmitted waveform proves to be an effective strategy in increasing the system's \ac{PLS}, even though the impact of an illegally deployed RIS is still non-negligible.
\begin{figure}[t]
\begin{center}
   \resizebox{0.45\textwidth}{!}{
\input{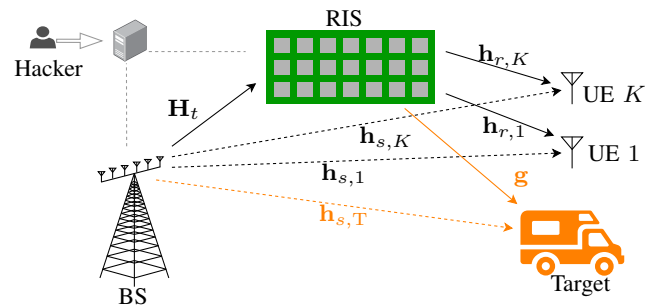}}
      \vspace{-4mm}
	 \caption{A RIS-aided ISAC network where a hacker has hacked into the RIS's control circuit.} \label{general scheme}
		\end{center}
  \vspace{-8mm}
\end{figure}
On the other hand, \cite{lin2023pain} assesses the impact of an active RIS, i.e., a RIS capable of amplifying the impinging signal, onto an Internet of Things network where the RIS aims at minimizing a single device's SNR. 
The authors analyze both the cases of known and unknown malicious RIS identity.
A significant challenge is that the attacks are difficult to detect because they are not carried out by generating interfering signals, as in traditional jamming. Instead, they involve configuring the RIS to degrade the system's performance. In \cite{rivetti2024malicious}, the impact of these silent attacks is demonstrated, further showing how destructive beamforming is susceptible to  \ac{CSI} uncertainties.
An important albeit often overlooked practical consideration about RIS is that individual elements, also known as \emph{pixels}, may fail: \cite{taghvaee2020error} gives an overview of the different types of errors that may arise in a practical RIS implementation and characterizes their impact on the RIS radiation pattern.
In \cite{zhang2022phase}, the authors propose a failure model to specify the amplitude and phase-shift of faulty elements and propose a diagnostic method while the authors of \cite{ozturk2024ris} assess the impact of RIS pixel failures onto downlink localization.

In this paper, we consider a \ac{MU}-\ac{MIMO} network equipped with an ISAC \ac{BS}, serving multiple \acp{UE} while sensing a target.
We first characterize the potential positive impact of a RIS deployment by devising a sensing-centric optimization problem computing active and passive beamforming vectors that maximize the sensing SNR under \acp{UE}' SINR constraints and a finite power budget.
Inspired by \cite{jiang2021intelligent}, this problem is solved through alternating optimization.
We then investigate the RIS destructive potential when, unbeknownst to the BS, the sensing SNR is minimized while the same minimum SINR is still guaranteed to the \acp{UE}, to make this attack harder to detect.
We further define a pixel fault model and assess its impact on both optimization algorithms.
Numerical simulations show that the RIS is equally capable of boosting and degrading the system performances and how the performances of both algorithms are hindered by the investigated pixel failures.

\emph{Notation}: $\odot$ represents the Hadamard (element-wise) product. Boldface lowercase and uppercase letters denote vectors and matrices.
The trace of the matrix $\XX$ is denoted by $\Tr(\XX)$.
 $\diag(\xx)$ represents the stacking of $\xx$ on the main diagonal of a matrix.
$[\hh]_{n:m}$ represents vector elements comprised between the $n$-th and $m$-th ones, denoted by $h_n$ and $h_m$, of vector $\hh$. 
The notation $\mathcal{CN}(0,\sigma^2)$ represents the zero-mean circularly symmetric complex Gaussian distribution with variance $\sigma^2$.

\section{System Model}
We consider the \ac{RIS}-assisted MU-\ac{MIMO} \ac{ISAC} network shown in Fig. \ref{general scheme}, where a \ac{BS}, equipped with $M$ transmit and $M$ receiving antennas and possessing digital beamforming capabilities, serves $K$ single-antenna \acp{UE} in the downlink while sensing the presence of a target.
The system is equipped with a \ac{RIS} made of $N$ reflective elements, whose phase-shifts are described by the vector $\ttheta=[\theta_1 \dots \theta_N]^\top$, where $|\theta_n|=1$ for $n=1,\dots, N$.
The static path between the \ac{BS} and \ac{UE} $k$  is denoted by $\hh_{\textrm{s},k}\in \mathbb{C}^{M}$ whereas the static path between the \ac{BS} and target is denoted by $\hh_{\textrm{s,T}}\in \mathbb{C}^{M}$.
The reflected paths are defined in a similar fashion, with $\hh_{\textrm{r},k} \in \mathbb{C}^{N}$ connecting the \ac{RIS} and  \ac{UE} $k$ while the target is connected to the \ac{RIS} via $\ggc \in \mathbb{C}^{N}$.
Lastly, the \ac{BS} is connected to the \ac{RIS} through the channel $\HH_t \in \mathbb{C}^{N\times M}$.
The end-to-end channel between the \ac{BS} and \ac{UE} $k$ is defined as  
\begin{align}
    \hh_k=\hh_{\textrm{s},k} + \HH_t^\hermit\underbrace{\diag(\ttheta)}_{\triangleq \TTheta}\hh_{\textrm{r},k} .
\end{align}
The received signal by \ac{UE} $k$ can be described by 
\vspace{-2mm}
\begin{align}
    y_k = \sum_{l=1}^L\hh_k^\hermit \ff_l t_l + w_k , 
\end{align}
where $\ff_l \in \mathbb{C}^{M}$ is the precoding vector associated with the $l$-th stream. We assume that the information meant for \ac{UE} $k$ is encoded into stream $k$, while an additional stream $K+1$ is used for sensing purposes. Hence, the total number of streams is $L=K+1$.
On the other hand, $t_l$ is the complex-valued unit-power information or sensing symbol and $w_k \sim \mathcal{CN}(0,\sigma_k^2)$ is the complex Gaussian receiver noise with variance $\sigma_k^2$.
As for the sensing performance, the two-way sensing channel is

\begin{align}
\HH_{\rm T} = c_r \HH_t^\hermit \TTheta \underbrace{\ggc\ggc^\hermit}_{\triangleq\TT} \TTheta^\hermit \HH_t + c_s \underbrace{\hh_{\textrm{s,T}}\hh_{\textrm{s,T}}^\hermit}_{\triangleq\hh_{\textrm{s,T}}}.
\end{align}
The \acp{RCS} associated with the two sensing paths, which also accounts for their pathloss, are modeled according to the Swerling-I model, that is, $c_r \sim \mathcal{CN}(0,\delta_r^2)$ and $c_s \sim \mathcal{CN}(0,\delta_s^2)$ and they take a single value throughout the transmission.
The monostatic sensing observation can be modeled as 
\vspace{-2mm}
\begin{align}
    \yy_{\rm T}= \HH_{\rm T}\FF\mathbf{t} + \mathbf{w}_{\rm T},
\end{align}
where $\FF=[\ff_1 \dots \ff_L]$ ,  $\mathbf{t}=[t_1 \dots t_L]^\top$, and the entries of $\mathbf{w}_{\rm T}$ are independent and distributed as $ \mathcal{CN}(0,\sigma_{\rm T}^2)$.
As for the RIS's pixel malfunctioning, we adopt the  \emph{clustered}-\emph{biased} model defined in \cite{taghvaee2020error}: the bias may be caused by bit-flip errors at the computing plane of the RIS or by external environmental factors, which can in turn cause a cascade effect onto the neighboring pixel's shifters, leading to the clustered nature of the failure.
We collect the neighboring faulty pixels in the set $\mathcal{Q}$. The affected pixels assume a phase-shift at a fixed distance $\kappa$ from the required one.
This is implemented by defining a failure mask $\mathbf{m}=[m_1 \dots m_N]^\top$ as~\cite{ozturk2024ris}
\vspace{-1mm}
\begin{align}
    m_n=\begin{cases}
     e^{j\kappa}, & \text{if}~n \in \mathcal{Q},  \\
     1, &  \text{otherwise} .
    \end{cases}.
\end{align}
The faulty RIS phase-shift vector is then obtained as $\ttheta^\text{fault}=\ttheta \odot \mathbf{m}$.

\section{Legitimate Sensing SNR Maximization}

The legitimate reason for deploying the \ac{RIS} is to improve the system performance. 
Hence, the \ac{BS} and \ac{RIS} jointly maximize
the sensing SNR while guaranteeing a minimum \ac{SINR} to the \acp{UE}. The choice of optimization function is motivated by the fact that the sensing SNR is expected to have a direct proportionality to various sensing tasks, such as detection and tracking. More specifically,\cite{cheng2023coordinated,cheng2024optimal} show that the detection probability monotonically increases with the sensing SNR.
The sensing \ac{SNR} can be defined as 
\begin{align} \label{eq:sensingSNR}
    \rho=\frac{\sum_{l=1}^L \rho_l}{\sigma_{\rm T}^2}  = \frac{\sum_{l=1}^L \mathbb{E}\left\{\norm{ \HH_{\rm T}\ff_l}^2 \right\} }{\sigma_{\rm T}^2} 
\end{align}
Similarly, the \ac{SINR} of the $k$-th \ac{UE} is described as 
\begin{align}
    \text{SINR}_k(\ttheta,\FF) = \frac{|\hh_k^\hermit \ff_k|^2}{\sigma^2_k + \sum_{\substack{l=1\\ l \neq k}}^L|\hh_k^\hermit \ff_l|^2}.
\end{align} 
The sensing SNR maximization problem, solved by the BS under the assumption of a collaborating \ac{RIS}, is the following:
\begin{subequations}
\begin{align}
     \underset{\ttheta,\{\ff_l\}_{l=1}^L}{\mathrm{maximize}} \,\, &~  \frac{\sum_{l=1}^L \rho_l}{\sigma_{\rm T}^2}\\
     \text{subject to} ~& \text{SINR}_k(\ttheta,\FF) \geq \gamma_k,\quad k=1,\dots, K, \\
     & \sum_{l=1}^L ||\ff_l||^2 \leq P, \\
     & |\theta_n| = 1, \quad n=1,\dots, N, 
\end{align}
\end{subequations}
where $P$ is the available transmit power of the \ac{BS}.
This problem is non-convex, mainly due to the coupling between the optimization variables and the non-convex unitary modulus constraints.
In this section, we will devise an \ac{AO} algorithm, which switches between optimizing the precoding vectors and RIS phase-shifts until convergence.
 
\subsection{Precoder optimization}\label{precoder subs}
In this subsection, we will consider the first subproblem of the \ac{AO} algorithm, which is the problem that retrieves the optimal precoder $\FF$ when $\TTheta$ is fixed.
It can be shown that the sensing SNR per stream, $\rho_l$, can be written as $\rho_l=\ff_l^\hermit\boldsymbol{\boldsymbol{\Gamma}}\ff_l$, where the matrix $\boldsymbol{\Gamma}$ is 
\vspace{-1.5mm}
\begin{align}  \nonumber
\boldsymbol{\Gamma} &= \delta_s^2\hh_{\textrm{s,T}}^\hermit\hh_{\textrm{s,T}} + \delta_m \hh_{\textrm{s,T}}^\hermit\HH_t^\hermit \TTheta\TT\TTheta^\hermit\HH_t\\ & \quad +\delta_m^* \HH_t^\hermit \TTheta\TT^\hermit\TTheta^\hermit\HH_t\hh_{\textrm{s,T}} \nonumber \\  & \quad +\delta_r^2\HH_t^\hermit \TTheta\TT^\hermit\TTheta^\hermit\HH_t\HH_t^\hermit\TTheta\TT\TTheta^\hermit\HH_t
\end{align}
and $\delta_m=\mathbb{E}[c_rc_s^*]$.
By defining the variable $\WW_l=\ff_l\ff_l^\hermit$, we obtain the following problem
\vspace{-2mm}
\begin{subequations}
\begin{align}
    \text{P}1 :=\underset{\WW_l\succeq 0, \forall l}{\mathrm{maximize}} \,\, &~  \frac{\sum_{l=1}^L \Tr\Big( \WW_l \boldsymbol{\Gamma}\Big)}{\sigma_\text{T}^2}\\
    \text{subject to} ~& \Tr(\WW_k \Tilde{\HH}_k) \geq \gamma_k \Bigg( \sum_{\substack{l \neq k}}\Tr(\WW_l \Tilde{\HH}_k)  \nonumber +\sigma^2_k \Bigg), \\   & ~k=1, \dots, K,  \\
     & \sum_{l=1}^L \Tr(\WW_l) \leq P, \\
     &\text{rank}(\WW_l) =1,~l=1, \dots, L,
\end{align}
\end{subequations}
where $\Tilde{\HH}_k=\hh_k\hh_k^\hermit$. We can apply \ac{SDR} to P1 by removing the rank one constraint. The resulting problem becomes convex and can be solved using general-purpose solvers like CVX.
During our simulations, we have observed that, for the scenario at hand, the optimal solutions $\{\WW_l^\text{opt}\}_{l=1}^K$ have rank one. Thus the relaxation is tight and $\{\ff_l^\text{opt}\}_{l=1}^K$ can be retrieved through an eigendecomposition.
We cannot say the same about $\WW_L$, thus, we employ the \emph{rand C} method from \cite{luo2004multivariate}, generating a set $\mathcal{Z}$ of randomized solutions that shall be denoted by $\ff_L^\text{rd}$.
We further define the set of feasible randomized solutions as  $\mathcal{S} = \{\ff_L^\text{rd}:\ff_L^\text{rd} \in \mathcal{Z},~ \Vert \ff_L^\text{rd} \Vert^2 \leq P - \sum_{k=1}^K \Vert \ff_k^\text{opt} \Vert^2 ,~ \text{SINR}_k(\ttheta,\Tilde{\FF}^\text{opt}) \geq \gamma_k,~k=1,\dots, K   \}  $, where $\Tilde{\FF}^\text{opt} = [\ff_1^\text{opt} \dots \, \ff_K^\text{opt} \ \ff_L^\text{rd}]$.
The sensing precoding vector is then obtained as
\begin{align}
    \ff_L^\text{opt} = \arg \max_{\ff_L^\text{rd} \in \mathcal{S}}~
    \Tr\Big( \ff_L^\text{rd}(\ff_L^\text{rd})^\hermit \boldsymbol{\Gamma}\Big)/\sigma_{\rm T}^2.
\end{align}

\vspace{-2mm}

\subsection{RIS phase-shift optimization}\label{RIS subs}
We now move on to the second part of the \ac{AO}, where we obtain $\TTheta$ while $\FF$ is fixed. 
We rewrite $\rho_l$, defined in \eqref{eq:sensingSNR}, as
\begin{align}
   \rho_l&= \delta_r^2\Tr\left(\HH_t\ff_l\ff_l^\hermit \HH_t^\hermit\TTheta\TT^\hermit\TTheta^\hermit\HH_t\HH_t^\hermit\TTheta\TT\TTheta^\hermit\right) \nonumber \\
&\quad+2\Re\left(\Tr\left(\delta_m^*\TTheta\TT^\hermit\TTheta^\hermit\HH_t\hh_{\textrm{s,T}}\ff_l\ff_l^\hermit\HH_t^\hermit\right)\right) \nonumber\\
&\quad + \delta_s^2\Tr\left(\hh_{\textrm{s,T}}^\hermit\hh_{\textrm{s,T}}\ff_l\ff_l^\hermit\right).
\end{align}
By utilizing the identity $\Tr(\AAb\XX\BB\XX)=(\vecc\left(\XX^\top\right)^\top)(\BB^\top \otimes \AAb) \vecc(\XX)$ and $\TT=\TT^\hermit$, the first term of $\rho_l$ can be rewritten as
\begin{align}
\vecc\left(\TTheta\TT\TTheta^\hermit\right)^\hermit \left(\left(\HH_t^*\HH_t^\top\right)\otimes\left(\delta_r^2\HH_\text{t} \ff_l\ff_l^\hermit\HH_t^\hermit\right)\right)\vecc\left(\TTheta\TT\TTheta^\hermit\right)
\end{align}
Then, thanks to the nature of $\TTheta$ \cite{jiang2021intelligent}, we can write that $\vecc(\TTheta\TT\TTheta^\hermit)^\hermit =\diag(\vecc(\TT))^\hermit\Tilde{\ttheta}$, where $\Tilde{\ttheta}$ collects the diagonal entries of $\TTheta^* \otimes \TTheta$.
We can finally rewrite the first term of $\rho_l$ as 
$\Tilde{\ttheta}^\hermit
    \QQ_l\Tilde{\ttheta}$, where 
\begin{align}
     \QQ_l &=\left(\diag\left(\text{vec}\left(\TT\right)\right)\right)^\hermit \left(\left(\HH_t^*\HH_t^\top\right)^\top\otimes\left(\delta_r^2\HH_\text{t} \ff_l\ff_l^\hermit\HH_t^\hermit\right)\right)\nonumber\\
     &\quad\times\diag\left(\text{vec}\left(\TT\right)\right).
\end{align}
We apply $\vecc(\TTheta\TT\TTheta^\hermit)^\hermit =\diag(\vecc(\TT))^\hermit\Tilde{\ttheta}$ to the second term of $\rho_l$ as well. Eventually, the sensing SNR per stream, excluding its constant terms, can be further rewritten as 
\begin{align}
    \rho_l=\Tilde{\ttheta}^\hermit
    \QQ_l\Tilde{\ttheta} + 2\Re\left(\Tilde{\ttheta}^\hermit\pp_l\right),
\end{align}
where 
\begin{align}
     \pp_l &= \diag(\vecc(\TT))^\hermit\vecc\left(\delta_m^*\HH_t\hh_{\textrm{s,T}}\ff_l\ff_l^\hermit\HH_t^\hermit\right).
\end{align}
The RIS phase-shift optimization problem under consideration is non-convex due to the unit-modulus constraints and the objective of maximizing a convex function. To circumvent this issue, we employ the \ac{MM} algorithm to maximize a lower bound on $\rho_l$ \cite{jiang2021intelligent}.
More specifically, we lower-bound $\Tilde{\ttheta}^\hermit
\QQ_l\Tilde{\ttheta}$ by computing its first-order Taylor expansion around the local point $\Tilde{\ttheta}^{(t)}$:
\begin{align}
     \Tilde{\ttheta}^\hermit
\QQ_l\Tilde{\ttheta} \geq 2\Re \left( \bb_l^{(t)\hermit} \Tilde{\ttheta}\right) - \Tilde{\ttheta}^{(t)\hermit}\QQ_l\Tilde{\ttheta}^{(t)},
\end{align}
where $\bb_l^{(t)}=\QQ_l\Tilde{\ttheta}^{(t)}$. Then, ignoring the constant terms, the objective function can be written as 
\begin{align}
 & \frac{1}{\sigma_{\rm  T}^2}  \sum_{l=1}^L2\left(\Re \left( \bb_l^{(t)\hermit} \Tilde{\ttheta} \right) +  \Re \left( \Tilde{\ttheta}^\hermit \pp_l \right)\right) \nonumber\\
& = \frac{1}{\sigma_{\rm  T}^2}  \sum_{l=1}^L
    2\Re\left (\ttheta^\hermit \left(\boldsymbol{\Sigma}_{\bb_l^{(t)}}+\boldsymbol{\Sigma}_{\pp_l}\right) \ttheta\right),
\end{align}
where $\boldsymbol{\Sigma}_{\bb_l^{(t)}} \in \mathbb{C}^{N \times N}$ is a matrix built by stacking the elements of $\bb_l^{(t)}$.
We define the  variable $\hat{\TTheta}=\hat{\ttheta}\hat{\ttheta}^\hermit$, where $\hat{\ttheta}=[\ttheta^\top,1 ]^\top$ and $\hh_{\textrm{r},k}=\diag(\hh_{\textrm{r},k})$. Then, the second subproblem of the \ac{AO} algorithm is
\begin{subequations}
\begin{align}
   \text{P}2 :=  \underset{\hat{\TTheta}\succeq0}{\textrm{maximize}}&~  \frac{1}{\sigma_{\rm T}^2}\sum_{l=1}^L 2\Tr\left( \boldsymbol{\Xi}_l^{(t)}\hat{\TTheta}\right) \\
     \text{subject to} ~& \left[\hat{\TTheta}\right]_{n,n} = 1, ~ n=1,\ldots, N+1, \\
     & \Tr\left( \RR_{k,k}\hat{\TTheta}\right) \geq \gamma_k \left( \sum_{l \neq k}\Tr\left( \RR_{k,l}\hat{\TTheta}\right) +     \sigma^2_k \right) \\
     &\text{rank}\left(\hat{\TTheta} \right)=1,
 \end{align}
 \end{subequations}
\begin{algorithm} [h!]
  \caption{Sensing SNR maximizing \ac{AO} algorithm}
  \begin{algorithmic}[1]
    \STATE \textbf{Initialize:} Randomly generate $\ttheta^{(0)}$, $t \leftarrow 0$
    \REPEAT
        \STATE Solve the relaxed version of P1 with $\ttheta= \ttheta^{(t)}$
        \STATE Retrieve $\FF^{(t+1)}$ through Gaussian randomization
         \STATE Solve the relaxed version of P2 with $\FF= \FF^{(t+1)}$
         \STATE Retrieve $\ttheta^{(t+1)}$ through Gaussian randomization
         \STATE $t \leftarrow t+1$
    \UNTIL{$\left|\rho^{(t)} - \rho^{(t-1)}\right| \leq \nu $}
    \STATE \textbf{Output:} $\ttheta^\text{opt},\FF^\text{opt}$
  \end{algorithmic}
\end{algorithm}

 where 
  \begin{align}
        \RR_{k,l} &= \begin{bmatrix}
          \hh_{\textrm{r},k}^\hermit \HH_t\ff_l\ff_l^\hermit\HH_t^\hermit \hh_{\textrm{r},k} & \hh_{\textrm{r},k}^\hermit\HH_t\ff_l\ff_l^\hermit\hh_{\textrm{s},k}  \\
          \hh_{\textrm{s},k}^\hermit \ff_l \ff_l^\hermit \HH_\text{t}^\hermit \hh_{\textrm{r},k}
           & |\hh_{\textrm{s},k}^{\hermit}\ff_l|^2
          \end{bmatrix},\\
          \boldsymbol{\Xi}_l^{(t)}&=\begin{bmatrix}
\boldsymbol{\Sigma}_{\bb_l^{(t)}}+\boldsymbol{\Sigma}_{\pp_l} & \mathbf{0}_{N \times 1} \\
          \mathbf{0}_{1 \times N} & 0
          \end{bmatrix}.
 \end{align}
By removing the rank-one constraints, 
the relaxed problem becomes convex. Hence, it can be solved by CVX, and we can retrieve a good sub-optimal solution via Gaussian randomization.
Following the \emph{rand C} method, we generate a set $\mathcal{R}$ of randomized solutions denoted by $\hat{\ttheta}^\text{rd}$. We normalize them as $\boldsymbol{\eta}(\hat{\ttheta}^\text{rd}) = \hat{\ttheta}^\text{rd} /\hat{\theta}^\text{rd}_{N+1}$ to ensure that the last element is one and then we define 
$\Tilde{\eta}(\hat{\ttheta}^\text{rd})_n = e^{j\angle\eta(\hat{\ttheta}^\text{rd})_n},~n=1,\dots, N$, to meet the unit modulus constraint.
We then define the set of feasible randomized solutions as  $\mathcal{E}= \Big\{\boldsymbol{\Tilde{\eta}}\left(\hat{\ttheta}^\text{rd}\right)   : \hat{\ttheta}^\text{rd}\in \mathcal{R},~ \text{SINR}_k\left(\boldsymbol{\Tilde{\eta}}\left(\hat{\ttheta}^\text{rd}\right) ,\FF^\text{opt}\right) \geq \gamma_k ,~k=1,\dots, K    \Big\} $, the  RIS phase-shifts are recovered as 
\begin{align}
    \hat{\ttheta}^\text{opt}=\arg \!\!\max_{\boldsymbol{\Tilde{\eta}}(\hat{\ttheta}^\text{rd}) 
    \in \mathcal{E}} \frac{2}{\sigma_{\rm T}^2}\sum_{l=1}^L\Tr\left( \boldsymbol{\Xi}_l^{(t)} \boldsymbol{\Tilde{\eta}}\left(\hat{\ttheta}^\text{rd}\right)
    \boldsymbol{\Tilde{\eta}}\left(\hat{\ttheta}^\text{rd}\right)^\hermit \right).
\end{align}
The steps of the \ac{AO} algorithm are outlined in Algorithm~1.

\section{Malicious RIS sensing SNR degradation}
We assume that the \ac{RIS} control circuit has been hacked. The hacker aims to degrade the sensing SNR so that the sensing target can go undetected.
To make the hacker's action hard to detect, the \acp{UE} are still guaranteed a minimum SINR.
Motivated by the collaboration between the BS and RIS, we assume that the hacker has access to the optimal precoders $\FF^\text{opt}$.
Additionally, we consider the worst-case scenario in which perfect \ac{CSI} is available to the hacker.
Hence, the hacker models the cascaded channel as $\hh_k = \hh_{\textrm{s},k} + \HH_t^\hermit\hh_{\textrm{r},k}\ttheta$.
According to these assumptions, the $\rho_l$ can be rewritten as
\begin{align}
\rho_l&=\delta_r^2\left|\ff_l^\hermit\HH_t^\hermit\GGc\ttheta\right|^2 \left\Vert \HH_t^\hermit\GGc\ttheta \right\Vert^2 \nonumber\\
& \quad +\ttheta^\hermit \MM_l\ttheta+\delta_s^2\ff_l^\hermit\hh_{\textrm{s,T}}^\hermit\hh_{\textrm{s,T}}\ff_l,
\end{align}
where 
\begin{align}
    \GGc &= \diag(\ggc),\\
    \MM_l&=\GGc^\hermit \HH_t \left(\delta_m^{*} \hh_{\textrm{s,T}}  \ff_l \ff_l^\hermit +\delta_m\ff_l\ff_l^\hermit \hh_{\textrm{s,T}}^\hermit \right)\HH_t^\hermit\GGc.
\end{align}
The sensing-degrading selection of $\TTheta$ is obtained by solving the  problem
\begin{subequations}
\begin{align}
   \text{P}3 :=  \underset{\ttheta}{\textrm{minimize}}&~  \frac{\sum_{l=1}^L \rho_l}{\sigma_{\rm T}^2} \\
     \text{subject to} ~& \text{SINR}_k(\ttheta,\FF^\text{opt}) \geq \gamma_k,~ k=1,\dots, K, \label{min sinr}\\
     & |\theta_n |=1,\quad n=1,\dots, N. \label{unit modulus}
 \end{align}
 \end{subequations}
 This problem is non-convex and, therefore, challenging to solve. However, we will demonstrate that each non-convex term of the objective function can be reformulated in a manageable manner, enabling us to solve P$3$ using the \ac{CCP} \cite{DiRenzo2020robust}. The first term of $\rho_l$ is the product between individually convex terms of $\ttheta$. We decouple those terms by introducing the variables $\mathbf{s}=[s_1 \dots s_L]^\top$ and $r$, redefining $\rho_l$ by removing the constant term as 
  $  \rho_l = \delta_r^2s_l^2r^2 +\ttheta^\hermit \MM_l\ttheta,$
and adding the constraints
\begin{align}
    &|\ff_l^\hermit\HH_t^\hermit\GGc\ttheta|^2 \leq s_l^2 
    \Vert \HH_t^\hermit\GGc\ttheta \Vert^2 \leq r^2 ,\quad l= 1,\dots, L\label{constr r original}.
\end{align}   
The product between $s_l^2$ and $r^2$ is still nonconvex, thus, we rewrite it by leveraging the square-completion method:  
\begin{align}\label{rho_l intermediate}
    \rho_l = \frac{\delta_r}{2}(s_l^2+r^2)^2-\frac{\delta_r}{2}s_l^4-\frac{\delta_r}{2}r^4+ \ttheta^\hermit \MM_l\ttheta.
\end{align}
The last term is a source of non-convexity because, albeit Hermitian symmetric, $\MM_l$ is not necessarily \ac{PSD}.
We then introduce the auxiliary variables $\boldsymbol{\alpha}=[\alpha_1 \dots \alpha_L]^\top$ and redefine $\rho_l$ as
\begin{align}
    \rho_l =& \frac{\delta_r}{2}(s_l^2+r^2)^2-\frac{\delta_r}{2}s_l^4-\frac{\delta_r}{2}r^4+\alpha_l
\end{align}
with the additional constraints
\begin{align}\label{quadratic original}\hat{\ttheta}^\hermit\underbrace{\begin{bmatrix}\MM_l & \boldsymbol{0}_{N\times 1} \\
        \boldsymbol{0}_{1 \times N} & 0
    \end{bmatrix}}_{\hat{\MM}_l}\hat{\ttheta} \leq \alpha_l,
\end{align}
where $\hat{\ttheta}=[\ttheta^\top,1]^\top$. 
We then apply the \ac{CCP} method \cite{DiRenzo2020robust} to \eqref{quadratic original} and obtain
\begin{align}\label{FPP constraint}
          &\hat{\ttheta}^\hermit \hat{\MM}_l^{(+)} \hat{\ttheta} +2\Re\left( \hat{\ttheta}^{(t)\hermit}\hat{\MM}_l^{(-)}\hat{\ttheta} \right)  \leq \alpha_l \nonumber \\
&+\hat{\ttheta}^{(t)\hermit}\hat{\MM}_l^{(-)}\hat{\ttheta}^{(t)}, \quad l=1, \dots, L,
 \end{align}
 where $\hat{\MM}_l^{(+)},\hat{\MM}_l^{(-)}$ are respectively the positive and negative semi-definite parts of $\hat{\MM}_l $ and $\hat{\ttheta}^{(t)}$ being a local point.
The second and third terms of \eqref{rho_l intermediate} are concave and are then locally upper bounded by their first-order Taylor expansion.
 We  define an approximated version of $\rho_l$  around the local point $(\mathbf{s}^{(t)},r^{(t)},\hat{\ttheta}^{(t)})$ by also removing the constant terms: 
\begin{align}
     \Tilde{\rho}_l^{(t)}=& \frac{\delta_r}{2}(s_l^2+r^2)^2-2\delta_rs^{(t),3}_ls_l-2\delta_r r^{(t),3}r+ \alpha_l.
 \end{align}
 The feasible-point pursuit successive convex approximation procedure \cite{mehanna2014feasible} is applied to \eqref{unit modulus},  decoupling it into
\begin{subequations}\label{ccp constr}
 \begin{align}
    &\left|\hat{\theta}_n\right|^2 \leq 1 + \xi_n~,~n=1,\ldots, N, \\
    &\left|\hat{\theta}_n^{(t)}\right|^2 - 2\Re\left( \hat{\theta}_n^*\hat{\theta}_n^{(t)}\right) \leq \xi_{N+n} -1 ~,~n=1,\ldots, N,
\end{align}   
\end{subequations}

\noindent where $\boldsymbol{\xi}=[\xi_1 \dots \xi_{2N}]^\top$ is a vector of auxiliary variables.
The minimum SINR constraint \eqref{min sinr} is written as
\begin{align}\label{CCP SINR}
    \gamma_k \left( \hat{\ttheta}^\hermit \UU_k \hat{\ttheta} + \sigma_k^2 \right)  
     - \hat{\ttheta}^{\hermit} \RR_{k,k}\hat{\ttheta} \leq 0,
\end{align}
where
\begin{align}
    \UU_k &= \begin{bmatrix}
        \hh_{\textrm{r},k}^\hermit \HH_t \FF_{-k}\FF_{-k}^\hermit \HH_t^\hermit \hh_{\textrm{r},k} &  \hh_{\textrm{r},k}^{\hermit} \HH_t \FF_{-k}\FF_{-k}^\hermit \hh_{\textrm{s},k} \\
        \hh_{\textrm{s},k}^\hermit \FF_{-k}\FF_{-k}^\hermit \HH_t^\hermit \hh_{\textrm{r},k} & \hh_{\textrm{s},k}^\hermit\FF_{-k}\FF_{-k}^\hermit\hh_{\textrm{s},k}
    \end{bmatrix}.
\end{align}
The notation $\FF_{-k}\in \mathbb{C}^{M \times L-1}$ indicates a matrix containing all the precoding vectors except the one assigned to the $k$-th UE. The constraint in
\eqref{CCP SINR} is then convexified as 
\begin{align}\label{SCA SINR}
     &\gamma_k \left( \hat{\ttheta}^\hermit \UU_k \hat{\ttheta} + \sigma_k^2 \right)- 2\Re \left( \hat{\ttheta}^{(t)\hermit} \RR_{k,k} \hat{\ttheta} \right)\nonumber \\
    & + \hat{\ttheta}^{(t)\hermit} \RR_{k,k} \hat{\ttheta}^{(t)}\leq v_k,\quad k=1, \dots, K ,
\end{align}
where $\mathbf{v}=[v_1 \dots v_K]^\top$ are auxiliary variables.
Lastly, \eqref{constr r original} are locally approximated as  
\begin{align}
    &|\boldsymbol{\tau}_l\hat{\ttheta}|^2 + s_l^{(t),2}-2s_l^{(t)}s_l \leq 0,~l=1,\ldots, L , \label{aux 1} \\&\Vert\hat{\HH}\hat{\ttheta}\Vert^2 + r^{(t),2}-2r^{(t)}r \leq 0, \label{aux 2}
\end{align}
where $\boldsymbol{\tau}_l=[\ff_l^\hermit \HH_t^\hermit\GGc, 0]$ and $\hat{\HH}=[\HH_t^\hermit\GGc, \mathbf{0}_{M \times 1}]$.
We are now finally able to define the convex reformulation of P$3$:
\begin{subequations}
 \begin{align}\label{minimization}
   \text{P}4:= \underset{\hat{\ttheta} , \mathbf{v} , \boldsymbol{\xi},\mathbf{s},r }{\textrm{minimize}}&~  \frac{\sum_{l=1}^L  \Tilde{\rho}_l^{(t)} }{\sigma_{\rm T}^2}+\lambda^{(t)}(||\boldsymbol{\xi}|| + ||\mathbf{v}||)\\
     \text{subject to} ~& 
    \hat{\theta}_{N+1}=1, \\
    &\eqref{FPP constraint},\eqref{ccp constr}, \eqref{SCA SINR},\eqref{aux 1},\eqref{aux 2},\\
    &\boldsymbol{\xi},\mathbf{v},\mathbf{s},r  \geq 0,
\end{align}
\end{subequations}
where  $\lambda^{(t)}$ is a penalty factor adjusting the impact of the auxiliary variables onto the optimization function.
The iterative procedure is described in Algorithm~\ref{alg 2}, with $\nu$ being a preset tolerance level.

 \begin{algorithm}[h!]
  \caption{Sensing SNR degradation algorithm}
  \label{alg 2}
  \begin{algorithmic}[1]
    \STATE \textbf{Initialize:} Set $t=0$ and $\lambda^{(0)}$, randomly generate $\ttheta^{(0)}$, $s_l^{(0)}= \left|\ff_l^\hermit\HH_t^\hermit\GGc\ttheta^{(0)}\right|^2 ,r^{(0)} = \left\Vert \HH_t^\hermit\GGc\ttheta^{(0)}\right\Vert^2$
    \REPEAT
        \STATE Retrieve $\left(\mathbf{s}^{(t+1)},r^{(t+1)},\hat{\ttheta}^{(t+1)}\right)$ by solving P4
        \STATE $\lambda^{(t+1)}=\min\left(\mu\lambda^{(t)},\lambda_\text{max} \right)$
        \STATE $t \leftarrow t+1$
    \UNTIL{$\left\Vert\hat{\ttheta}^{(t)} - \hat{\ttheta}^{(t-1)}\right\Vert \leq \nu, \Vert \boldsymbol{\xi}\Vert \leq \nu, \Vert \mathbf{v}\Vert \leq \nu$}
    \STATE \textbf{Output:} $\ttheta^\text{opt}$
  \end{algorithmic}
\end{algorithm}

\section{Numerical Results}

In this section, we will demonstrate the ability of the RIS to either improve or degrade the system performance.
 The simulation results are averaged over $500$ independent channel realizations with $\left| \mathcal{Z}\right| = \left| \mathcal{R}\right|=1000 $.
The number of \ac{BS} antennas is $M=30$ and the number of \acp{UE} is $K=4$. Unless otherwise specified, $N=15$, with a number of faulty pixels equal to $\left| \mathcal{Q}\right|=4$.
The static paths $\hh_{\textrm{s},k}$, $\hh_{\textrm{s,T}}$, and $\ggc$ are independent and identically distributed (i.i.d.) Rician fading with a K-factor of $10$, whereas $\HH_t$ is a scattering-free \ac{LoS} channel.
Lastly, $\hh_{\textrm{r},k}$ are i.i.d.\ Rayleigh fading.
The static and reflected path \ac{RCS} variances are set to $\delta_r^2=\delta_s^2=10^{-5}$ whereas $\delta_m=9\cdot10^{-6}$.
We assume an equal minimum \ac{SINR} for all \acp{UE}, denoted by $\gamma$ and it is equal to $2$, unless otherwise specified.
The \ac{BS} available power is $3$\,dBW and the \acp{UE} and sensing noise variances are set to $\sigma_k^2 = \sigma_{\rm T}^2=1$.
We consider the $2$D area shown in Fig. \ref{simulation scenario}, where the \acp{UE} are randomly distributed within the highlighted area.

\begin{figure}[t!]
    \centering
     \resizebox{0.45\textwidth}{!}{\includegraphics{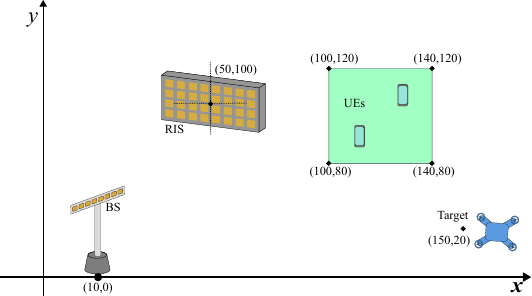}}
     \vspace{-2mm}
    \caption{Illustration of the simulation setup.}
     \label{simulation scenario}
\end{figure}
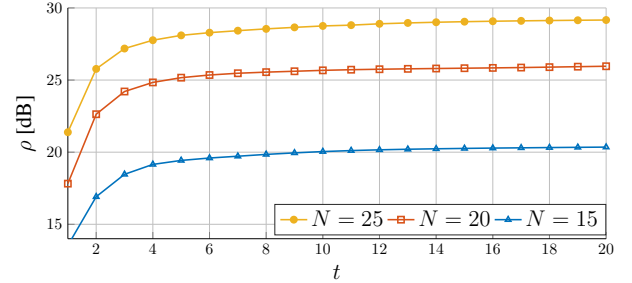
\begin{figure}[t!]
\centering
\resizebox{0.45\textwidth}{!}{
%
%
\definecolor{mycolor1}{rgb}{0.00000,0.44700,0.74100}%
\definecolor{mycolor2}{rgb}{0.85000,0.32500,0.09800}%
\definecolor{mycolor3}{rgb}{0.92900,0.69400,0.12500}%
\usetikzlibrary{spy}
\begin{tikzpicture}[spy using outlines={rectangle, magnification=5,connect spies}]

\begin{axis}[%
width=4.721in,
height=2.02566in,
at={(0.758in,0.481in)},
scale only axis,
xmin=1,
xmax=20,
xlabel style={font=\color{white!15!black}\Large },
xlabel={$t$},
ymin=14,
ymax=30,
ylabel style={font=\color{white!15!black}\Large},
ylabel={$\rho$ [dB]},
axis background/.style={fill=white},
axis x line*=bottom,
axis y line*=left,
legend columns =3,
xmajorgrids,
ymajorgrids,
legend style={at={(1,0.15)},legend cell align=left, align=left, draw=white!15!black, font=\Large}
]
\addplot [color=mycolor3, line width=1.0pt,,mark=*,mark options={solid}]
  table[row sep=crcr]{%
1	21.3844663653145\\
2	25.7741820272514\\
3	27.185305700083\\
4	27.7626222462423\\
5	28.0996121736981\\
6	28.2868858332746\\
7	28.4235586894906\\
8	28.5486785757202\\
9	28.6536985828223\\
10	28.7517835395533\\
11	28.8075421786011\\
12	28.9001678974966\\
13	28.9602109417152\\
14	29.008406235441\\
15	29.0473522344268\\
16	29.0785445566459\\
17	29.1050368521491\\
18	29.1271534768999\\
19	29.1460571059465\\
20	29.1625346793132\\
};
\addlegendentry{$N=25$}

\addplot [color=mycolor2, line width=1.0pt,mark=square,mark options={solid}]
  table[row sep=crcr]{%
1	17.8199711440571\\
2	22.6311982442063\\
3	24.2060170614847\\
4	24.8369759175886\\
5	25.1661299033734\\
6	25.3499661775102\\
7	25.4722599172946\\
8	25.5494747097549\\
9	25.6100597742154\\
10	25.6729775312695\\
11	25.7183521505656\\
12	25.7508367505491\\
13	25.7778716586577\\
14	25.8003899197459\\
15	25.8221967432986\\
16	25.8432756148553\\
17	25.8668841660269\\
18	25.8962343826595\\
19	25.9278756568702\\
20	25.957267873152\\
};
\addlegendentry{$N=20$}

\addplot [color=mycolor1, line width=1.0pt,mark=triangle,mark options={solid}]
  table[row sep=crcr]{%
1	13.5354271867519\\
2	16.9116654403304\\
3	18.4647480176534\\
4	19.1523197088495\\
5	19.4285318975964\\
6	19.5950942349931\\
7	19.7196750855644\\
8	19.8462422329314\\
9	19.9507852762801\\
10	20.0427333636593\\
11	20.1111570971269\\
12	20.1639056328959\\
13	20.2039456758194\\
14	20.2354240412474\\
15	20.2612204360739\\
16	20.2839927718631\\
17	20.3021090116942\\
18	20.3202510387536\\
19	20.3368502002593\\
20	20.3512350612259\\
};
\addlegendentry{$N=15$}

\end{axis}
\end{tikzpicture}%
}
\vspace{-2mm}
     \caption{Alg.~1 $\rho$ maximization vs 
 $t$ for different values of $N$.}
     \label{legitimate}
 \end{figure}

\begin{figure}[t!]
\centering
\resizebox{0.45\textwidth}{!}{
%
%
\definecolor{mycolor1}{rgb}{0.00000,0.44700,0.74100}%
\definecolor{mycolor2}{rgb}{0.85000,0.32500,0.09800}%
\definecolor{mycolor3}{rgb}{0.92900,0.69400,0.12500}%
\definecolor{mycolor5}{rgb}{0.4940, 0.1840, 0.5560}%
\begin{tikzpicture}

\begin{axis}[%
width=4.721in,
height=2.01566in,
at={(0.758in,0.481in)},
scale only axis,
xmin=1,
xmax=20,
xlabel style={font=\color{white!15!black}\Large },
xlabel={$t$},
ymin=10,
ymax=22,
ylabel style={font=\color{white!15!black}\Large},
ylabel={$\rho$ [dB]},
axis background/.style={fill=white},
axis x line*=bottom,
axis y line*=left,
legend columns =1,
xmajorgrids,
ymajorgrids,
legend style={at={(1,0.5)},legend cell align=left, align=left, draw=white!15!black, font=\Large}
]
\addplot [color=mycolor5, line width=1.0pt]
  table[row sep=crcr]{%
1	13.5354271867519\\
2	16.9116654403304\\
3	18.4647480176534\\
4	19.1523197088495\\
5	19.4285318975964\\
6	19.5950942349931\\
7	19.7196750855644\\
8	19.8462422329314\\
9	19.9507852762801\\
10	20.0427333636593\\
11	20.1111570971269\\
12	20.1639056328959\\
13	20.2039456758194\\
14	20.2354240412474\\
15	20.2612204360739\\
16	20.2839927718631\\
17	20.3021090116942\\
18	20.3202510387536\\
19	20.3368502002593\\
20	20.3512350612259\\
};
\addlegendentry{not faulty}
\addplot [color=mycolor1, line width=1.0pt,mark=triangle,mark options={solid}]
  table[row sep=crcr]{%
1	11.5476482336029\\
2	15.3472662689613\\
3	17.1251756476995\\
4	17.9893660973642\\
5	18.3160646770812\\
6	18.4857361100662\\
7	18.5982418061471\\
8	18.6796389176951\\
9	18.7509657980936\\
10	18.8002298192046\\
11	18.842237589138\\
12	18.8661903919368\\
13	18.8914093185028\\
14	18.9099685881385\\
15	18.9247945821692\\
16	18.9369051900766\\
17	18.9470705718436\\
18	18.9590791191195\\
19	18.9691555083463\\
20	18.9770303062947\\
};
\addlegendentry{$\kappa=\pi/6$}

\addplot [color=mycolor2, line width=1.0pt,mark=*,mark options={solid}]
  table[row sep=crcr]{%
1	11.0638431556543\\
2	14.8323929543101\\
3	16.6201933419756\\
4	17.4958810682997\\
5	17.8276383081912\\
6	18.0007135530956\\
7	18.1159330773832\\
8	18.1996513723712\\
9	18.2729816610871\\
10	18.3237874132552\\
11	18.3664863878539\\
12	18.391608890272\\
13	18.4175524184054\\
14	18.4372117602512\\
15	18.4525820950241\\
16	18.4653594824341\\
17	18.475940182424\\
18	18.4886663839675\\
19	18.4991485228758\\
20	18.5071703685979\\
};
\addlegendentry{$\kappa=\pi/4$}

\addplot [color=mycolor3, line width=1.0pt,mark=square,mark options={solid}]
  table[row sep=crcr]{%
1	10.3635867372387\\
2	14.0961639961451\\
3	15.8960346786176\\
4	16.7845541630835\\
5	17.121195103854\\
6	17.2973967082243\\
7	17.4149837219458\\
8	17.5008742835009\\
9	17.5760828653267\\
10	17.6284395611047\\
11	17.6717582036518\\
12	17.6982058846654\\
13	17.7248255745046\\
14	17.7456635545835\\
15	17.7615742541392\\
16	17.775086027307\\
17	17.7861083238967\\
18	17.7996140082948\\
19	17.8104983640199\\
20	17.818654425849\\
};
\addlegendentry{$\kappa=\pi/3$}

\end{axis}
\end{tikzpicture}%
}
     \caption{Alg.~1 $\rho$ maximization vs $t$ in presence of failures.}
     \label{legitimate faulty}
     \vspace{-5mm}
 \end{figure}
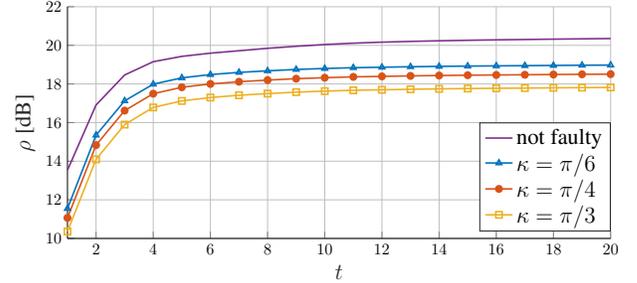

\begin{figure}[t!]
\centering
\resizebox{0.45\textwidth}{!}{
%
%
\definecolor{mycolor1}{rgb}{0.00000,0.44700,0.74100}%
\definecolor{mycolor2}{rgb}{0.85000,0.32500,0.09800}%
\definecolor{mycolor3}{rgb}{0.92900,0.69400,0.12500}%
\definecolor{mycolor5}{rgb}{0.4940, 0.1840, 0.5560}%

\begin{tikzpicture}[spy using outlines={rectangle, magnification=5,connect spies}]

\begin{axis}[%
width=4.521in,
 height=1.91566in,
at={(0.758in,0.481in)},
scale only axis,
xmin=0,
xmax=15,
xlabel style={font=\color{white!15!black}\Large},
xlabel={$t$},
ymin=-28,
ymax=23,
ylabel style={font=\color{white!15!black}\Large},
ylabel={$\rho$ [dB]},
axis background/.style={fill=white},
axis x line*=bottom,
axis y line*=left,
xmajorgrids,
ymajorgrids,
legend style={at={(1,0.6)},legend cell align=left, align=left, draw=white!15!black, font=\Large}
]
\addplot [color=mycolor1, line width=1.0pt,mark=triangle,mark options={solid}]
  table[row sep=crcr]{%
0	20.3512350612259\\
1	-19.0909742388625\\
2	-19.901177553508\\
3	-20.6792522745201\\
4	-21.074952189253\\
5	-21.31576798982\\
6	-21.4343971207191\\
7	-21.532013174406\\
8	-21.5615511438736\\
9	-21.5173888463332\\
10	-21.575232361849\\
11	-21.5877496724961\\
12	-21.5828176194497\\
13	-21.5602010908665\\
14	-21.5843248648436\\
15	-21.611555921998\\
};
\addlegendentry{$\gamma=2$}

\addplot [color=mycolor2, line width=1pt,mark=square,mark options={solid}]
  table[row sep=crcr]{%
0	20.3512350612259\\
1	-19.2051652452021\\
2	-20.2671375914841\\
3	-21.2668839676259\\
4	-21.7974545804327\\
5	-22.1799706431892\\
6	-22.4681819865164\\
7	-22.6386815388103\\
8	-22.7519056877684\\
9	-22.831832714751\\
10	-22.8893097429176\\
11	-22.9326203395084\\
12	-22.960544077548\\
13	-22.9730271878295\\
14	-22.9939187140465\\
15	-23.0100434323731\\
};
\addlegendentry{$\gamma=1$}

\addplot [color=mycolor3, line width=1pt,mark=*,mark options={solid}]
  table[row sep=crcr]{%
0	20.3512350612259\\
1	-20.9713611090405\\
2	-22.8731491804407\\
3	-23.2706766432197\\
4	-23.7139450253272\\
5	-23.9321306658283\\
6	-21.4838275493661\\
7	-23.5637994827045\\
8	-23.985082645414\\
9	-23.9512360729153\\
10	-23.9564039566349\\
11	-23.9971045881675\\
12	-24.044484404744\\
13	-24.0795917335433\\
14	-24.0997810274049\\
15	-24.11290227957\\
};
\addlegendentry{$\gamma=0.5$}

\coordinate (spypoint) at (axis cs:13,-22.5);
\coordinate (spyviewer) at (axis cs:4
,1);
\spy[width=3cm,height=2cm] on (spypoint) in node [fill=white] at (spyviewer);

\end{axis}
\end{tikzpicture}%
}
     \caption{Alg.~2 $\rho$ minimization vs $t$ for different values of $\gamma$.}
     \label{hacker}
     \vspace{-5mm}
 \end{figure}
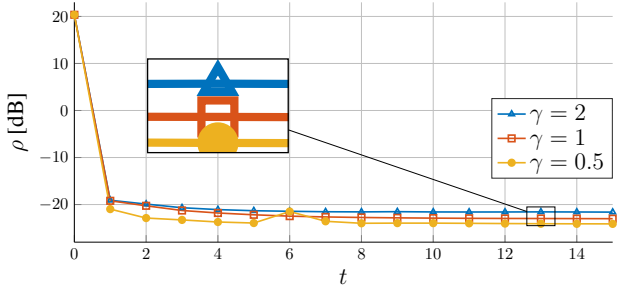

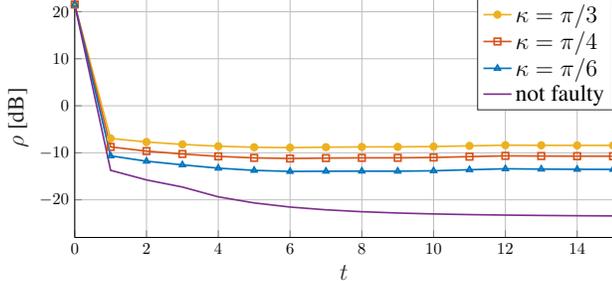
\begin{figure}[t!]
\centering
\resizebox{0.45\textwidth}{!}{
%
%
\definecolor{mycolor1}{rgb}{0.00000,0.44700,0.74100}%
\definecolor{mycolor2}{rgb}{0.85000,0.32500,0.09800}%
\definecolor{mycolor3}{rgb}{0.92900,0.69400,0.12500}%
\definecolor{mycolor5}{rgb}{0.4940, 0.1840, 0.5560}%
\begin{tikzpicture}

\begin{axis}[%
width=4.521in,
 height=2.01566in,
at={(0.758in,0.481in)},
scale only axis,
xmin=0,
xmax=15,
xlabel style={font=\color{white!15!black}\Large},
xlabel={$t$},
ymin=-28,
ymax=23,
ylabel style={font=\color{white!15!black}\Large},
ylabel={$\rho$ [dB]},
axis background/.style={fill=white},
axis x line*=bottom,
axis y line*=left,
xmajorgrids,
ymajorgrids,
legend style={at={(1,1)},legend cell align=left, align=left, draw=white!15!black, font=\Large}
]
\addplot [color=mycolor3, line width=1.0pt,mark=*,mark options={solid}]
  table[row sep=crcr]{%
0	21.534325498387\\
1	-6.94361057887473\\
2	-7.70086581197734\\
3	-8.21411229730922\\
4	-8.59842932516605\\
5	-8.82254168936919\\
6	-8.90518978216438\\
7	-8.80590347632289\\
8	-8.75106153800353\\
9	-8.7352776778315\\
10	-8.6698720102066\\
11	-8.52277996443392\\
12	-8.38417610654967\\
13	-8.40998479708074\\
14	-8.4243256433684\\
15	-8.42984431053126\\
};
\addlegendentry{$\kappa=\pi/3$}
\addplot [color=mycolor2, line width=1.0pt,mark=square,mark options={solid}]
  table[row sep=crcr]{%
0	21.534325498387\\
1	-8.75091129112306\\
2	-9.64038192908278\\
3	-10.2521107482333\\
4	-10.7615181733057\\
5	-11.0774826506681\\
6	-11.2086627487925\\
7	-11.1229265710289\\
8	-11.0772776828486\\
9	-11.0693734028118\\
10	-10.9951237906602\\
11	-10.8150271168596\\
12	-10.6498076365965\\
13	-10.6913978922527\\
14	-10.7153711664677\\
15	-10.7255549028569\\
};
\addlegendentry{$\kappa=\pi/4$}
\addplot [color=mycolor1, line width=1.0pt,mark=triangle,mark options={solid}]
  table[row sep=crcr]{%
0	21.534325498387\\
1	-10.6502310099002\\
2	-11.7877500525999\\
3	-12.5626448657949\\
4	-13.2673111348284\\
5	-13.7357030200818\\
6	-13.9576465959529\\
7	-13.9179253278522\\
8	-13.8990192776728\\
9	-13.9103413603583\\
10	-13.828379410316\\
11	-13.6016361984903\\
12	-13.3998611740217\\
13	-13.4698646796572\\
14	-13.5110895874142\\
15	-13.5295468090134\\
};
\addlegendentry{$\kappa=\pi/6$}
\addplot [color=mycolor5, line width=1.0pt]
  table[row sep=crcr]{%
0	21.534325498387\\
1	-13.7222661726692\\
2	-15.7684309130773\\
3	-17.2781379859359\\
4	-19.363616811301\\
5	-20.6602652795105\\
6	-21.5382520244036\\
7	-22.1244901185678\\
8	-22.5290107754833\\
9	-22.8076652795173\\
10	-23.0040306921921\\
11	-23.145116859995\\
12	-23.2480326083796\\
13	-23.3243090691457\\
14	-23.3814611971143\\
15	-23.4247657658117\\
};
\addlegendentry{not faulty}
\end{axis}
\end{tikzpicture}%
}
\vspace{-2mm}
     \caption{Alg.~2 $\rho$ minimization vs. $t$ in the presence of failures. }
     \label{hacker faulty}
     \vspace{-5mm}
 \end{figure}

\subsection{Sensing SNR maximization}
We will first test the sensing SNR maximization capabilities of Algorithm~1. Fig. \ref{legitimate} shows that $\rho$ is a monotonically increasing function of the number of iterations $t$ and convergence is reached in $20$ iterations.
A larger RIS size allows us to reflect more energy towards the target, resulting in a higher $\rho$.
There are diminishing returns in increasing $N$. When $N$ goes from $15$ to $20$, the final value of $\rho$ increases by $5$\,dB, whereas when $N$ goes from $20$ to $25$ the increase is halved.
Fig. \ref{legitimate} shows the impact of clustered-biased pixel failures \cite{taghvaee2020error} onto Algorithm~1. This kind of failure does not alter the maximization trend, but leads to convergence to a lower $\rho$-value.
For example,  a big phase bias magnitude of $\pi/3$ attains a $\rho$ value only $2$\,dB lower than the one obtained with a perfectly functioning RIS, showing that nonideal hardware can be tolerated by the system.
\subsection{Malicious RIS action}
 Fig. \ref{hacker} shows how Algorithm~2 causes SNR reduction. One iteration is sufficient to obtain a $40$\,dB  SNR reduction w.r.t. to the maximized value obtained in the previous section.
This reduction further increases in the span of 15 iterations, reaching a final value around $-22$\,dB.
Such an attack is complicated to detect as the communication SINRs are untouched, thus the communication \acp{UE} will not see any difference in their service.
Reducing $\gamma$ would improve the SNR reduction at the expense of a higher chance of the attack being discovered. However, a $50\%$ reduction in $\gamma$ corresponds to a mere $2$\,dB SNR reduction.
We attribute the bump at iteration 6 of the SNR curve with $\gamma=0,5$ to an outlying realization among the averaged ones: it is worth mentioning that said bump does not hinder the SNR convergence process and $\rho$ at iteration $15$ is lower than $\rho$ at iteration $6$.
Fig.~\ref{hacker faulty} assesses the impact of the considered pixel failures model onto Algorithm~2. Contrary to what we observed in Fig.~\ref{legitimate faulty}, faulty RIS hardware has a bigger impact on the final sensing SNR.
A bias of $\pi/6$ is sufficient to induce a $10$\,dB penalty onto the final value of $\rho$, making this algorithm less tolerant towards faulty hardware.

\section{Conclusions}
We have investigated the potential benefits and harm of having a RIS in an ISAC MU-MIMO network.
The positive side of the RIS's presence has been investigated by devising an optimization algorithm maximizing the sensing target SNR under the communication SINR constraints and a finite power budget.
The negative side arises when a hacker takes control of the RIS, computing its phase-shifts to minimize the target SNR, under the same SINR constraints. The hacker can make the sensing target practically invisible to the system, which is a major concern if ISAC is meant to be used for surveillance.
We have further considered the presence of clustered pixel failures in the RIS.
Numerical simulations have exemplified how great both potentials are and how the investigated pixel failures caused the final SNR value to be at a fixed offset from the one obtained with an unimpaired RIS.

 \bibliographystyle{IEEEtran}
\bibliography{IEEEabrv,auxiliary/biblio}
\end{document}